\documentclass[onecolumn,amsmath,amssymb,superscriptaddress,nofootinbib,11pt,a4paper]{revtex4}

\pdfoutput=1
\usepackage[T1]{fontenc}
\usepackage{CJK}
\usepackage{xcolor}
\usepackage{amsfonts}
\usepackage{epstopdf}
\usepackage{wrapfig} 
\usepackage{subcaption}
\usepackage{graphicx}  
\usepackage{dcolumn}   
\usepackage{bm}
\usepackage{float}
\usepackage[T1]{fontenc}
\usepackage{CJK}
\usepackage{xcolor}
\usepackage{amsfonts}
\usepackage{epstopdf}
\usepackage{wrapfig} 
\usepackage{subcaption} 
\usepackage{graphicx}  
\usepackage{dcolumn}   
\usepackage{bm}
\usepackage{float}
\usepackage[utf8]{inputenc}
\usepackage[T1]{fontenc}

\begin{document}

\title{Hotspot Image Driven by Magnetic Reconnection in Kerr-anti-de Sitter Black Holes}

\author{Xiao-Xiong Zeng}
\email{xxzengphysics@163.com}
\affiliation{College of Physics and Electronic Engineering, Chongqing Normal University, \\Chongqing 401331, China}

\author{Ke Wang\footnote{Electronic address: kkwwang2025@163.com  (Corresponding author)}}
\affiliation{School of Material Science and Engineering, Chongqing Jiaotong University, \\Chongqing 400074, China}
\begin{abstract}
{Based on the Comisso-Asenjo mechanism, we investigate the kinematic images of plasma before and after magnetic reconnection in Kerr-Anti-de Sitter(Kerr-AdS) black holes. Following a brief review of the Comisso-Asenjo process in Kerr-AdS black holes, we introduce the hotspot model and the imaging method. Building upon these foundational theories, we obtain the trajectory of the plasma and the temporal evolution of the hotspot images. It is found that there are three flares within the observing time, which is driven by the Comisso-Asenjo mechanism. We also discuss the influence of the cosmological parameter on the hotspot imaging. The results indicate that the hotspot image enlarges as the absolute value of $\Lambda$ increases, demonstrating that the cosmological constant significantly affects the hotspot.}
\end{abstract}

\maketitle
 \newpage
\section{Introdution}
In recent years, a growing body of observational evidence has confirmed the existence of black holes \cite{5,6}. The Gravity collaboration has also reported near-infrared flare events from the vicinity of the event horizon of Sgr A* \cite{7,8}. Currently, numerous studies focus on black hole images and flares. Regarding flares, an effective approach is to investigate hotspots orbiting the central black hole \cite{12,13,14,15,16,17,18}. This method typically employs a semi-analytic hotspot model and uses numerical methods to compute photon trajectories. Research on hotspot imaging and related issues is highly important and necessary, as it aids in understanding the observed black hole flare phenomena in astronomy.

Recently, the study of magnetic reconnection processes within the framework of General Relativity has developed rapidly \cite{19,20,21,22}. Ref. \cite{19} introduced the magnetic reconnection process in curved spacetime, namely the Sweet-Parker model, but its reconnection rate is too slow. Subsequently, Ref. \cite{21} extended it to the Petschek model. By introducing a slow shock structure, faster reconnection was achieved. Magnetic reconnection events can accelerate magnetofluids to near the speed of light. When such events occur within the ergosphere of a black hole, they can serve as an ignition mechanism for the Penrose process \cite{23}, thereby enabling the extraction of the black hole's rotational energy. The energy extraction mechanism of this process was elucidated by Comisso and Asenjo \cite{1}. Utilizing the Comisso-Asenjo process for energy extraction has been widely extended to other rotating black holes \cite{24,25,26,27,28,29,38}. However, there is very little research on the imaging of hot spots in the Penrose process driven by magnetic reconnection. Ref. \cite{33} studied the hotspot imaging of this process in Kerr black holes, the observation of the plasmoid motion from this process, and found that it could potentially produce three flares, with the first flare potentially serving as a distinctive signature of energy extraction via the Penrose process. Ref. \cite{34} extended this work to Kerr-Sen spacetime, not only finding similar phenomena but also discovering that the expansion parameter has a significant impact on the hotspot imaging. In this paper, we aim to further investigate in anti-de Sitter(AdS) spacetime with a cosmological constant whether the magnetic reconnection process can drive hotspot imaging, and if so, what influence the cosmological parameter has on this process.

AdS spacetime is a vacuum solution of Einstein's field equations with a constant negative curvature \cite{30}. Its geometric structure can be analogized to an infinite four-dimensional "saddle-shaped" surface. Its most notable characteristic is the presence of a finite, accessible timelike boundary, enabling physical phenomena within the spacetime to be entirely described by a theory residing on this boundary. This holographic nature makes it a cornerstone of modern theoretical physics, particularly through the AdS/CFT correspondence \cite{31,32}. This duality establishes an equivalence between a theory of quantum gravity in AdS spacetime and a conformal field theory without gravity on its boundary, thereby providing a powerful theoretical laboratory for investigating black holes, quantum gravity, and strongly interacting systems. This paper investigates the hotspot image driven by magnetic reconnection in Kerr-AdS black holes. The results indicate that, besides the presence of similar flares, the hotspot image enlarges as the absolute value of $\Lambda$ increases, demonstrating a significant influence of the cosmological constant on the hotspot image.

The remainder of this paper is organized as follows. In Section 2, we provide a brief description of the magnetic reconnection process in Kerr-AdS black holes. Section 3 gives a concise introduction to the hotspot model and the imaging method. Our numerical results are presented in Section 4. A summary is given in Section 5.

\section{Magnetic Reconnection Process in Kerr-AdS Black Holes}
The particle motion for the hotspot imaging in this work is based on the Comisso-Asenjo mechanism. Therefore, we begin by reviewing the magnetic reconnection process described in Ref. \cite{1}. Throughout this paper, we use natural units $(c=G=1)$.

The line element of the Kerr-AdS metric in Boyer-Lindquist (BL) coordinates is given by \cite{2,3}
\begin{equation}
\mathrm{d}s^{2} = -\frac{\Delta_{r}}{I^{2}\rho^{2}}(\mathrm{d}t-a\sin^{2}\theta \mathrm{d}\phi)^{2}+\frac{\Delta_{\theta}\sin^{2}\theta}{I^{2}\rho^{2}}\left[a\mathrm{ d}t-(r^{2}+a^{2})\,\mathrm{d}\phi\right]^{2}+\frac{\rho^{2}}{\Delta_{r}}\mathrm{d}r^{2}+\frac{\rho^{2}}{\Delta _{\theta}}\mathrm{d}\theta^{2},
\end{equation}
where
\begin{equation}
\rho^{2}=r^{2}+a^{2}\cos^{2}\theta,\Delta_{\theta} = 1+\frac{1}{3}\Lambda a^{2}\cos^{2}\theta,I = 1+\frac{1}{3}\Lambda a^{2},\Delta_{r} = -\frac{1}{3}\Lambda r^{2}\left(r^{2}+a^{2}\right)+r^{2}-2Mr+a^{2}.
\end{equation}
Here, $a$ is the spin, $M$ is the black hole mass, and $\Lambda$ is the cosmological constant. Spacetime with $\Lambda<0$ is Kerr-AdS, while with $\Lambda>0$ it is Kerr-de Sitter(Kerr-dS). The case $\Lambda=0$ corresponds to Kerr spacetime. In this paper, we adopt units where the mass $M=1$ and consider Kerr-AdS spacetime with $\Lambda<0$. The black hole horizon is determined by $\Delta_{r}=0$. This equation has two positive real roots; the larger root is the event horizon, and the smaller root is the Cauchy horizon. The boundary of the ergosphere is determined by $g_{tt}=0$. This equation has one root outside the event horizon, which defines the ergosphere boundary. Both the event horizon and the ergosphere boundary are defined by quartic equations. Their analytical expressions are quite lengthy, so in this paper, we directly utilize numerical results. Following Ref. \cite{1}, we will consider a simplified approximation where the current sheet moves on a stable Keplerian circular orbit in the equatorial plane ($\theta=\pi/2$). In Kerr-AdS black holes, the Keplerian angular velocity for massive particles on circular orbits is given by \cite{4}
\begin{equation}
\Omega= \frac{\sqrt{1-\frac{\Lambda M^2 r^3}{3}}}{a \sqrt{1-\frac{\Lambda M^2 r^3}{3}}\pm r^{1.5}}.
\end{equation}
Here, $+$ and $-$ represent corotating and counterrotating orbits, respectively. These orbits can extend from infinity down to the photon sphere. As energy extraction requires the current sheet to be located within the ergosphere, and since the counterrotating photon sphere lies outside it, only corotating orbits are possible. Therefore, in this paper, we will only consider corotating orbits. The spacetime line element can be expressed in the 3+1 form as
\begin{equation}
ds^{2}=g_{\mu\nu}dx^{\mu}dx^{\nu}=-\alpha^{2}dt^{2}+\sum_{i=1}^{3}\bigl {(}h_idx^{i}-\alpha\beta^{i}dt\bigr{)}^{2},   
\end{equation}
where
\begin{equation}
\alpha=\sqrt{-g_{tt}+\frac{g_{t\phi}^{2}}{g_{\phi\phi}}}\,,\beta^{i}=\delta_{i\phi}\frac{h_\phi\,\omega^{\phi}} {\alpha},\omega^{\phi}=-g_{t\phi}/g_{\phi\phi},h_i=\sqrt{g_{ii}}. 
\end{equation}
In the above equation, the specific forms of the corresponding metric components are
\begin{equation}
\begin{aligned}
g_{tt} &= \frac{ - \Delta_{r} + a^{2} \Delta_{\theta} \sin^{2}\theta }{ I^{2} \rho^{2} } \\
g_{t\phi}&= \frac{  a \sin^{2}\theta \left( \Delta_{r} - (r^{2} + a^{2}) \Delta_{\theta} \right) }{ I^{2} \rho^{2} } \\
g_{\phi\phi}& = \frac{ \sin^{2}\theta \left[ \Delta_{\theta} (r^{2} + a^{2})^{2} - a^{2} \sin^{2}\theta \Delta_{r} \right] }{ I^{2} \rho^{2} }.
\end{aligned}
\end{equation}
In the Zero Angular Momentum Observer(ZAMO) reference frame, the tetrad is given by
\begin{equation}
\hat{e}_t = {\alpha}^{-1} (\partial_0 + \omega^\phi \partial_3), \hat{e}_r = \frac{1}{h_r} \partial_1,  \hat{e}_\theta = \frac{1}{h_\theta} \partial_2,\hat{e}_\phi = \frac{1}{h_\phi} \partial_3. 
\end{equation}
Therefore, the transformation relation for the four-velocity between the BL reference frame and the ZAMO reference frame is
\begin{equation}
\hat{U}^{\mu}=\hat{\gamma}\left\{1, \hat{v}^{(r)}, 0, \hat{v}^{(\phi)}\right\}=\left\{\frac{E-\omega^{\phi} L}{\alpha}, h_r U^{r}, 0, \frac{L}{h_\phi}\right\}                 ,\label{7}
\end{equation}
where
\begin{equation}
\hat{v}=\sqrt{\left(\hat{v}^{(r)}\right)^{2}+\left(\hat{v}^{(\phi)}\right)^{2}},\hat{\gamma}=\frac{1}{\sqrt{1 - \hat{v}^{2}}}. 
\end{equation}
The Keplerian velocity in the ZAMO reference frame is
\begin{equation}
\hat{v}_{K}=\frac{1}{\alpha}h_\phi \Omega-\beta^{\phi}. 
\end{equation}
Then the four-velocity of the current sheet is
\begin{equation}
\hat{u}_{K} = (\hat{\gamma}_{K},0,0,\hat{\gamma}_{K}\hat{v}_{K}).    
\end{equation}
According to \eqref{7}, the energy and angular momentum of the current sheet in the BL frame are
\begin{equation}
L=h_\phi\hat{\gamma}_{K}\hat{v}_{K},E=\omega^\phi L+\alpha\hat{\gamma}_{K}.\label{11}
\end{equation}
To satisfy the circular orbit condition, the $E$ and $L$ in \eqref{11} must be real numbers. The tetrad of the fluid rest frame is established as
\begin{equation}
\sigma_{t} = \hat{\gamma}_{K}(\hat{e}_{t} + \hat{v}_{K}\hat{e}_{\phi}),
\sigma_{r} = \hat{e}_{r},
\sigma_{\theta} = \hat{e}_{\theta},
\sigma_{\phi} = \hat{\gamma}_{K}(\hat{v}_{K}\hat{e}_{t} + \hat{e}_{\phi}). 
\end{equation}
Assuming the azimuthal angle of the magnetic field lines in the fluid rest frame is $\xi$, the three-velocity of the plasma ejected from the reconnection layer in the fluid rest frame is
\begin{equation}
v^{\prime\mu} = \pm v_{out}(\cos\xi\sigma_{\phi}^{\mu} + \sin\xi\sigma_{r}^{\mu}),   
\end{equation}
where $+$ and $-$ correspond to the accelerated and decelerated plasmas, respectively, and $v_{out}$ denotes the magnitude of the outflow velocity observed in the fluid rest frame. The four-velocity of the plasma ejected from the reconnection layer in the fluid rest frame is then given by
\begin{equation}
u^{\prime\mu} = \gamma_{out}[\sigma_{t}^{\mu} \pm v_{out}(\cos\xi\sigma_{\phi}^{\mu} + \sin\xi\sigma_{r}^{\mu})],   
\end{equation}
where $\gamma_{out}$ is the Lorentz factor corresponding to $v_{out}$, satisfying \cite{1}
\begin{equation}
v_{ out }=\frac{\sqrt{\sigma}}{\sqrt{1+\sigma}},  \gamma_{out} = \frac{1}{\sqrt{1 - v_{out}^{2}}},
\end{equation}
where $\sigma$ is the upstream magnetization parameter, satisfying
\begin{equation}
\sigma=B_0^2/w,  
\end{equation}
where $w$ is the enthalpy density and $B_0$ is the magnetic field. In the ZAMO reference frame, the four-velocity of the plasma ejected from the reconnection layer is
\begin{equation}
\hat{u}^{\mu} = \gamma_{out}[\hat{\gamma}_{K}(\hat{e}_{t}^{\mu} + \hat{v}_{K}\hat{e}_{\phi}^{\mu}) \pm v_{out}(\cos\xi(\hat{\gamma}_{K}(\hat{v}_{K}\hat{e}_{t}^{\mu} + \hat{e}_{\phi}^{\mu})) + \sin\xi\hat{e}_{r}^{\mu})] = (\hat{\gamma}_{out},\hat{\gamma}_{out}\hat{v}^{r},0,\hat{\gamma}_{out}\hat{v}^{\phi}),    
\end{equation}
where
\begin{equation}
\hat{v}^{\phi} = \frac{\hat{v}_{K} \pm v_{out}\cos\xi}{1 \pm \hat{v}_{K}v_{out}\cos\xi},
\hat{v}^{r} = \pm \frac{v_{out}\sin\xi}{\hat{\gamma}_{K}\pm \hat{\gamma}_{K}\hat{v}_{K}v_{out}\cos\xi},
\hat{\gamma}_{out} = \gamma_{out}\hat{\gamma}_{K}(1 \pm \hat{v}_{K}v_{out}\cos\xi), \label{aaa}
\end{equation}
According to \eqref{7}, the energy, angular momentum, and radial momentum of the accelerated and decelerated plasmas in the BL frame are
\begin{equation}
L^\prime=h_\phi\hat{\gamma}_{out}\hat{v}^{\phi}=p_\phi,E^\prime=\alpha\hat{\gamma}_{out}+\omega^\phi L^\prime=-p_t, p_r= \hat{\gamma}_{out}\hat{v}^{r}  h_r.\label{18}
\end{equation}
\eqref{18} gives the equations of motion with lowered indices for the ejected plasma. Since the magnetic reconnection process is highly efficient and can rapidly convert magnetic energy into plasma kinetic energy, we adopt the single-fluid approximation and neglect the electromagnetic field tensor part. The energy-momentum tensor of the fluid is obtained as
\begin{equation}
T^{\mu\nu}=pg^{\mu\nu}+w\mathcal{U}^{\mu}\mathcal{U}^{\nu},    
\end{equation}
where $p$ is the proper pressure, and $\mathcal{U}$ is the four-velocity of the plasma ejected from the reconnection layer in the BL frame. The energy at infinity is given by
\begin{equation}
e^{\infty}=-\alpha g_{\mu 0}T^{\mu 0}.
\end{equation}
Substituting the metric components and the energy-momentum tensor yields
\begin{equation}
e^{\infty}=\alpha w\hat{\gamma}_{out}^{2}-\alpha p+\alpha\beta^{\phi}w\hat{\gamma}_{out}^{2}\hat{v}^{\phi}.   
\end{equation}
Assuming the plasma is treated as an adiabatic incompressible fluid, then
\begin{equation}
e^{\infty}=\alpha w\hat{\gamma}_{out}-\alpha p/\hat{\gamma}_{out}+\alpha\beta^{\phi}w\hat{\gamma}_{out}\hat{v}^{\phi}.  \label{bbb}    
\end{equation}
Substitute \eqref{aaa} into \eqref{bbb}, the infinity energy per unit enthalpy of the plasma ejected from the reconnection layer can be approximated as \cite{1}
\begin{equation}
\varepsilon_{\pm}=e^{\infty}/w=\alpha \hat{\gamma}_{K}\left[\left(1+\beta^{\phi} \hat{v}_{K}\right)\sqrt{1+\sigma} \pm \cos \xi\left(\beta^{\phi}+\hat{v}_{K}\right) \sqrt{\sigma} - \frac{1}{4} \frac{\sqrt{1+\sigma} \mp \cos{\xi} \hat{v}_{K} \sqrt{\sigma}}{\hat{\gamma}_{K}^{2}\left(1+\sigma-\cos ^{2}{\xi} \hat{v}_{K}^{2} \sigma\right)}\right].\label{19} 
\end{equation}
For energy extraction, the following two conditions must be satisfied \cite{1}
\begin{align}
\varepsilon_{-}<0\,\,\text{and} \,\,\varepsilon_{+}>0.
\end{align}
where we assume the plasma is relativistically hot, hence the relation between its enthalpy density and proper pressure is $w=4p$, with the adopted polytropic index being $\Gamma = 4/3$.

\section{Hotspot Model and Imaging Technique}
This section focuses on the hotspot model and the imaging technique. As in  Ref. \cite{33}, our goal is not to analyze the impact of the radiation mechanism on imaging or flares, thus the model excludes specifics of the hotspot's emission spectrum. The hotspot emission is assumed to be isotropic and frequency-independent, idealizing it as a broadband source with a flat spectrum. The plasmoid is modeled as a transparent hotspot whose emissivity adheres to a Gaussian distribution
\begin{equation}
J_{\nu_{s}}=e^{\frac{-l^{2}}{2s^{2}}}\,, \label{27} 
\end{equation}
where $J_{\nu_{s}}$ denotes the emissivity, $s$ represents the width of the Gaussian distribution, and $l$ is the distance from the center of the hotspot. Imaging the moving hotspot requires tracing the source's trajectory and computing the associated radiative transfer. The trajectory of the light source is determined via numerical integration of the Hamilton-Jacobi form of the geodesic equation, as given by \eqref{18}. For handling radiative transfer, images are generated using a numerical backward ray-tracing technique incorporating a fisheye camera model, see Appendix B of Ref. \cite{9} for details. The camera is defined within the ZAMO reference frame, with its tetrad given by
\begin{equation}
\hat{e}_{(0)}=\frac{g_{\phi\phi}\,\partial_{0}-g_{\phi t}\,\partial_{3}}{\sqrt{g_{\phi\phi}\Big(g_{\phi t}^{2}-g_{\phi\phi}g_{tt}\Big)}},\hat{e }_{(1)}=-\frac{\partial_{1}}{h_r}, \hat{e}_{(2)}=\frac{\partial_{2}}{h_\theta}\,, \hat{e}_{(3)}=-\frac{\partial_{3}}{h_\phi}.   
\end{equation}
The intensity on the image plane is given by the radiative transfer equation\cite{10,Zeng:2025kqw,Zeng:2021dlj,Zeng:2021mok}
\begin{equation}
\frac{d\mathcal{I}}{d\lambda} = \frac{J_{\nu}}{\nu^{2}} , \quad \text{where} \quad \mathcal{I} = \frac{I_{\nu}}{\nu^{3}}.\label{29}
\end{equation}
Here, $\lambda$ is the affine parameter along the null geodesic, $I_{\nu}$ is the specific intensity, and $J_{\nu}$ is the emissivity at frequency $\nu$. Absorption is neglected in the model. $I_{\nu}$ is obtained by integrating Eq. \eqref{29} after substituting the emissivity from Eq. \eqref{27}. Our numerical results comprise a time series of snapshots. The photon arrival time at the image plane is defined as the orbital time plus the light propagation time from the hotspot to the observer. The flux centroid position on the camera plane is also computed for each frame. The flux projected onto the $(i,j)$-th screen pixel, using the camera model from Ref. \cite{9}, is calculated as \cite{11}
\begin{equation}
F(i,j)=I_{\nu}(i,j)S\cos\left[2\arctan\left(\frac{1}{n}\tan\left(\frac{\alpha_{ f}}{2}\right)\sqrt{\left(i-\frac{n+1}{2}\right)^{2}+\left(j-\frac{n+1}{2}\right) ^{2}}\right)\right],\label{30}  
\end{equation}
where $S$ is the area per pixel, $n$ is the total pixel count along each axis (with $i,j$ from 1 to $n$), and $\alpha_{f}$ is the camera field of view. The centroid position $\vec{x}_{c}(t)$ for each image is then computed as
\begin{equation}
\vec{x}_{c}(t)=\frac{\displaystyle\sum_{i,j}\vec{x}(i,j)F(i,j)}{\displaystyle\sum_{i,j}F(i,j)}\,,   \label{31} 
\end{equation}
with $\vec{x}(i,j)$ being the spatial coordinates of pixel $(i,j)$. The total snapshot flux, $\sum_{i,j}F(i,j)$, gives the flux pertinent to $\vec{x}_{c}(t)$. Finally, Eqs. \eqref{30} and \eqref{31} allow for the tracking of the brightness centroid position and its associated flux over time.

\section{Numerical Results}
\begin{figure}[!h]
  \centering
  \begin{subfigure}{0.22\textwidth}
    \centering
    \includegraphics[width=\linewidth]{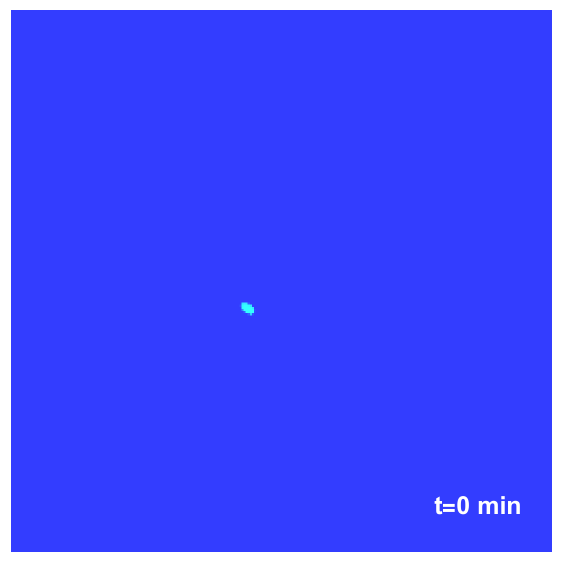} 
  \end{subfigure}
  \begin{subfigure}{0.22\textwidth}
    \centering
    \includegraphics[width=\linewidth]{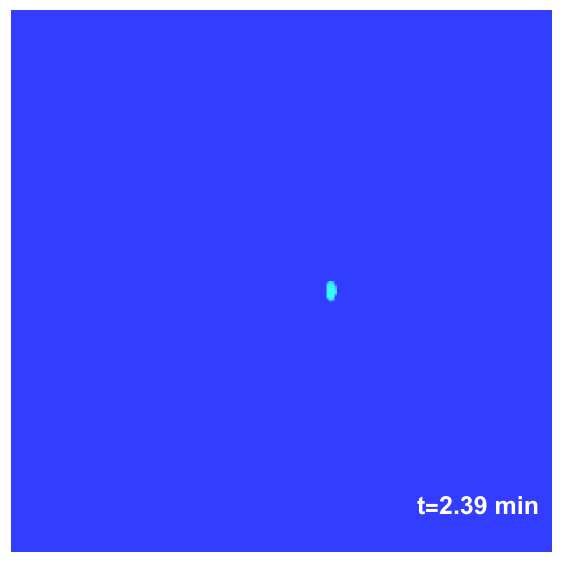} 
  \end{subfigure}
 \begin{subfigure}{0.22\textwidth}
    \centering
    \includegraphics[width=\linewidth]{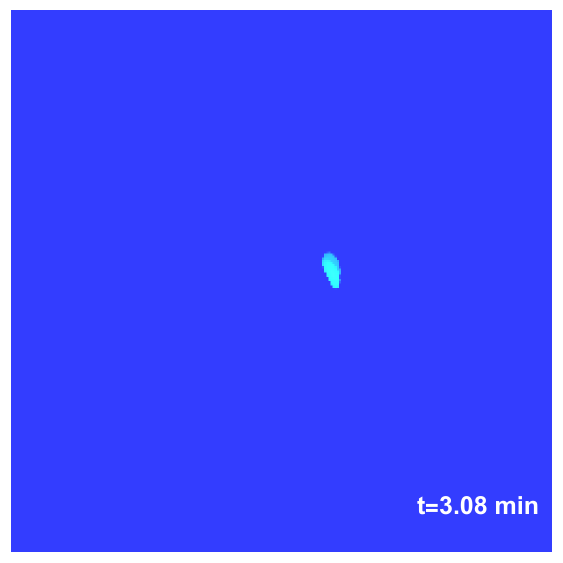}  
  \end{subfigure}
 \begin{subfigure}{0.22\textwidth}
    \centering
    \includegraphics[width=\linewidth]{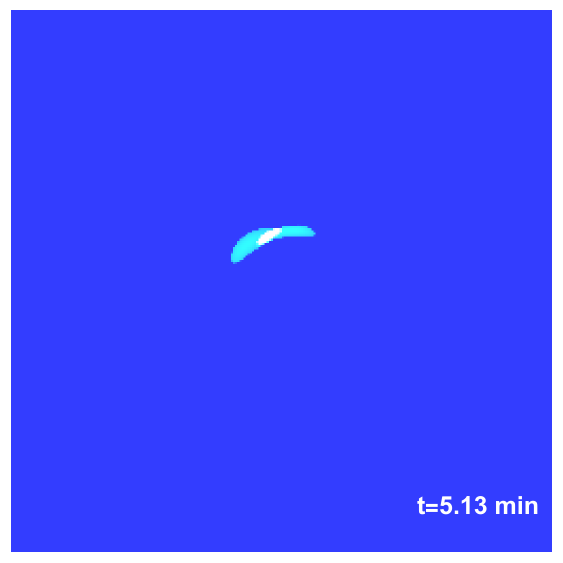} 
  \end{subfigure}
 \begin{subfigure}{0.22\textwidth}
    \centering
    \includegraphics[width=\linewidth]{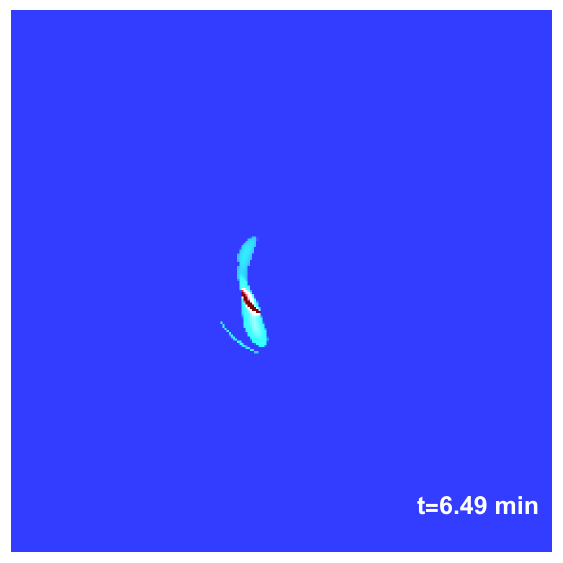}  
  \end{subfigure}
 \begin{subfigure}{0.22\textwidth}
    \centering
    \includegraphics[width=\linewidth]{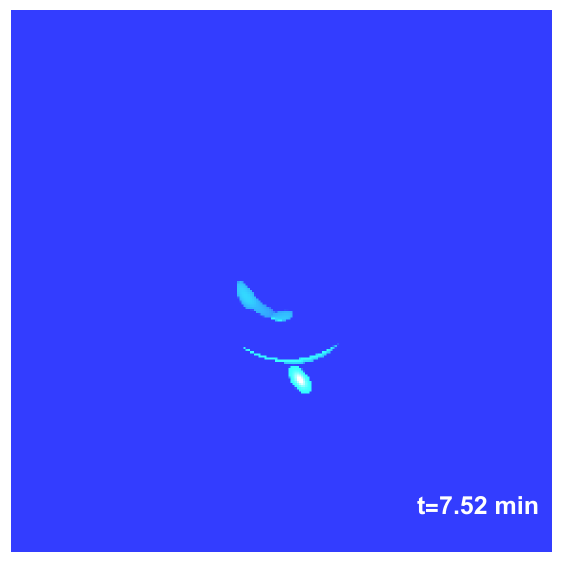} 
  \end{subfigure} 
 \begin{subfigure}{0.22\textwidth}
    \centering
    \includegraphics[width=\linewidth]{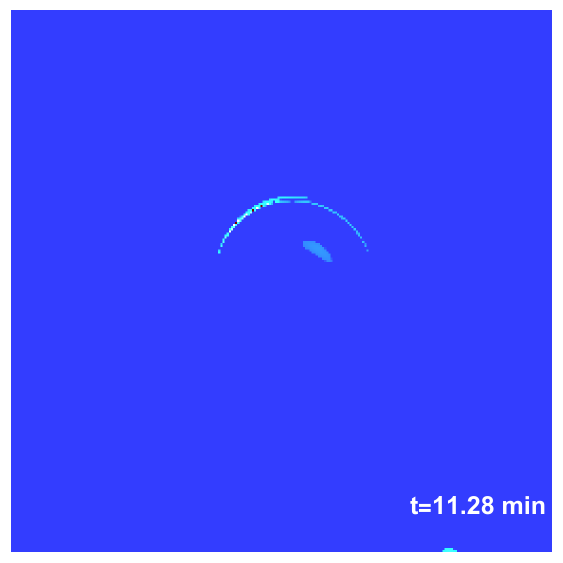} 
  \end{subfigure} 
  \begin{subfigure}{0.22\textwidth}
    \centering
    \includegraphics[width=\linewidth]{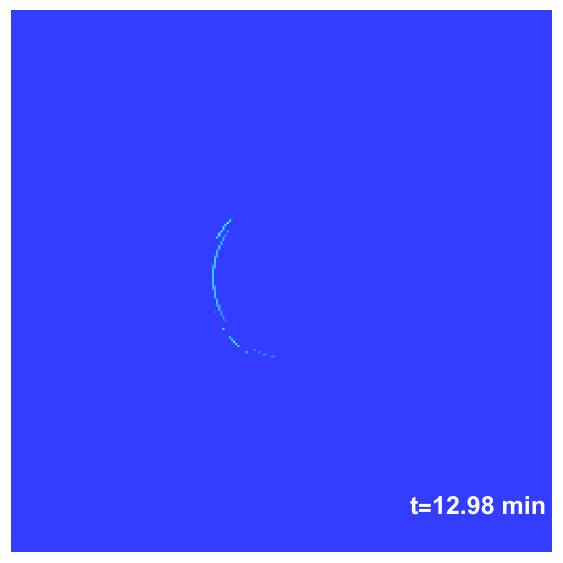} 
  \end{subfigure} 
  \caption{Time evolution of the plasma hotspot distribution for $\Lambda=-0.00002$.}
  \label{fig:1}
\end{figure}

\begin{figure}[!h]
  \centering
  \begin{subfigure}{0.26\textwidth}
    \centering
    \includegraphics[width=\linewidth]{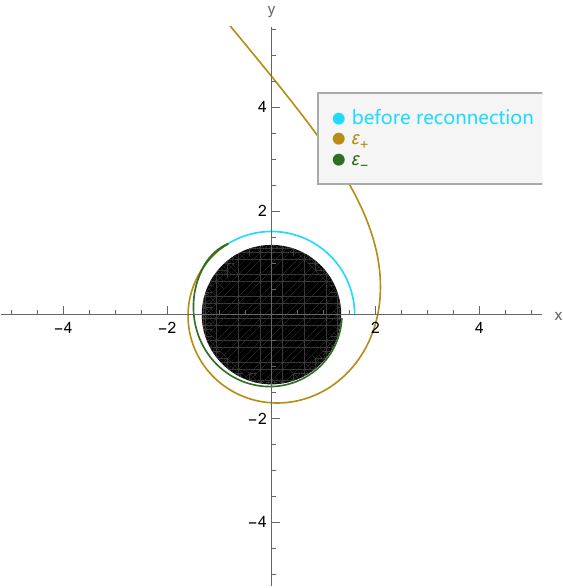}
    \caption{}
  \end{subfigure}
  \begin{subfigure}{0.26\textwidth}
    \centering
    \includegraphics[width=\linewidth]{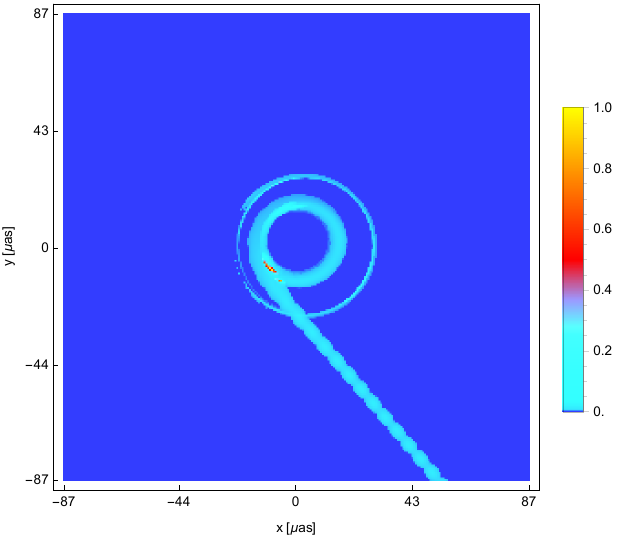}
    \caption{} 
  \end{subfigure}
  \begin{subfigure}{0.26\textwidth}
    \centering
    \includegraphics[width=\linewidth]{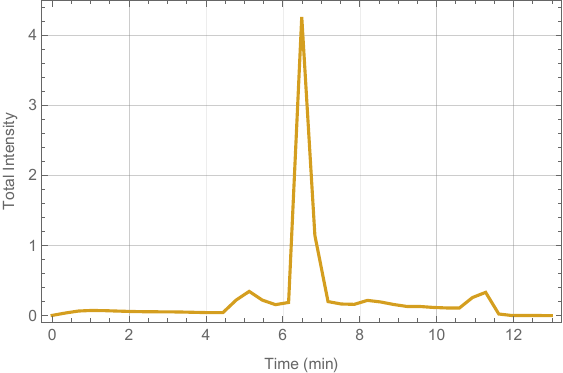}
    \caption{}
    \label{fig:2c}
  \end{subfigure}
  \begin{subfigure}{0.26\textwidth}
    \centering
    \includegraphics[width=\linewidth]{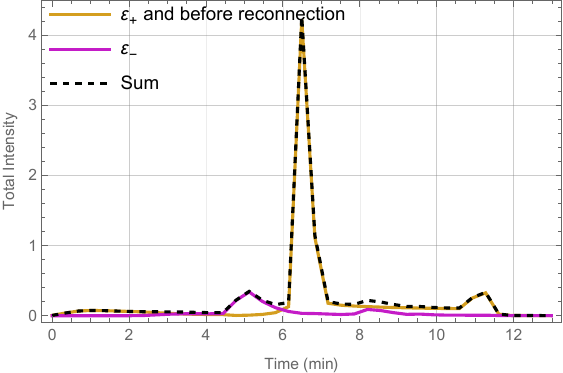} 
    \caption{}  
     \label{fig:2d}
  \end{subfigure}
  \begin{subfigure}{0.26\textwidth}
    \centering
    \includegraphics[width=\linewidth]{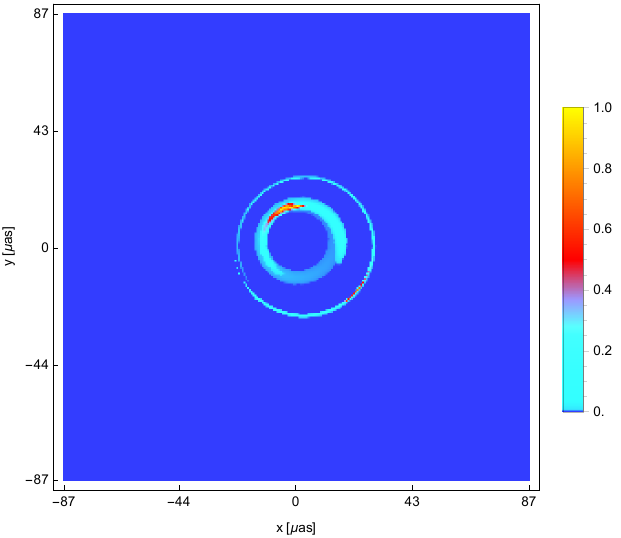} 
    \caption{} 
  \end{subfigure}
\caption{$\Lambda=-0.00002$. (a) Trajectory of the plasmoid in a two-dimensional Cartesian coordinate system. Cyan represents the state before magnetic reconnection, yellow represents the accelerated plasma, green represents the decelerated plasma, and black represents the black hole. (b) Normalized intensity distribution of the hotspot as seen by the observer, showing the time-averaged radiation intensity on the observation plane. The intensity values are normalized by $I/I_{MAX}$. The outer ring in the figure represents secondary or higher-order images. (c) Hotspot emission light curve, showing the total flux versus observed time. (d) The solid yellow line represents the light curve for $\varepsilon_{+}$ and the case before magnetic reconnection, the solid purple line represents the light curve produced solely by $\varepsilon_{-}$, and the black dashed line represents the total observed light curve. (e) Normalized hotspot intensity distribution produced solely by the $\varepsilon_{-}$ plasma.}
  \label{fig:2}
\end{figure}

We present the numerical observational results in this section. The parameters are fixed as $\sigma=25$, $\xi=\pi/20$, $a=0.94$, $s=0.2$, $\Lambda=-0.00002$, with the observer at $(\phi_0, \theta_0, r_{\text{obs}})=(\pi/2, \pi/10, 200)$. The dominant $X$-point is at $r_X=1.6$. As established, equatorial current sheets fragment due to plasmoid instability, forming multiple $X$-points; the dominant one, at the separatrix intersection governing global reconnection, is considered here. The initial observation time on the screen is $t=0$. The current sheet lies within the ergosphere $r\in(1.34113,1.99993)$ and satisfies the circular orbit condition. From Eq. \eqref{19}, $\varepsilon_{-}=-0.50356$ and $\varepsilon_{+}=9.89102$, meeting the energy extraction criteria. Results are shown in Figs. \ref{fig:1} and \ref{fig:2}.

Figure \ref{fig:1} shows the temporal evolution, beginning at $t=0$ with a Keplerian hotspot appearing on the screen. The instantaneous magnetic reconnection process concludes at $t=2.39$ min. By $t=3.08$ min, the hotspots of the accelerated and decelerated plasmas become resolvable. The hotspot reaches the first flare position at $t=5.13$ min, and the second flare at $t=6.49$ min, where secondary or higher-order images also become visible. A clear separation is observed at $t=7.52$ min, the accelerated plasma moves outward towards infinity, while the decelerated plasma lingers near the event horizon. At $t=11.28$ min, the hotspot coincides with the third flare; the decelerated plasma is on the verge of crossing the event horizon, and the accelerated plasma has reached the screen boundary, poised to exit. Finally, by $t=12.98$ min, both plasma components have exited the screen, leaving behind only the secondary or higher-order images. This image is similar to that in Ref. \cite{33}.

Analysis of the total light curve in Fig. \ref{fig:2c} reveals three flares within the observation window, characterized by a weak, a bright, and again a weak flare, consistent with Ref. \cite{33}. Fig. \ref{fig:2d} indicates that the first weak flare is generated by the decelerated plasmoid, while the subsequent two flares originate from the accelerated plasmoid. Furthermore, $\varepsilon_{-}$ produces two bumps, as noted previously, with the second being fainter than the first. This second bump is not classified as a distinct flare due to its minimal contribution to the total light curve. The brightest flare arises because the accelerated plasmoid acquires significant kinetic energy and undergoes substantial Doppler blueshift, as explained in Ref. \cite{33}. The third flare stems from the Doppler blueshift of the secondary image of the accelerated plasmoid, whereas the first flare from the decelerated plasmoid comes from the Doppler blueshift of its primary image. For the detailed mechanism of flare production, see Appendix A of Ref. \cite{33}. Ref. \cite{33} identified the first flare from $\varepsilon_-<0$ as unique and a potential signature of ongoing Penrose process energy extraction, albeit with limitations such as reduced prominence at high spins. Our hotspot imaging in Kerr-AdS spacetime also shows this first flare from $\varepsilon_-$, demonstrating the value and significance of this investigation in this spacetime.

\begin{figure}[!h]
  \centering
  \begin{subfigure}{0.24\textwidth}
    \centering
    \includegraphics[width=\linewidth]{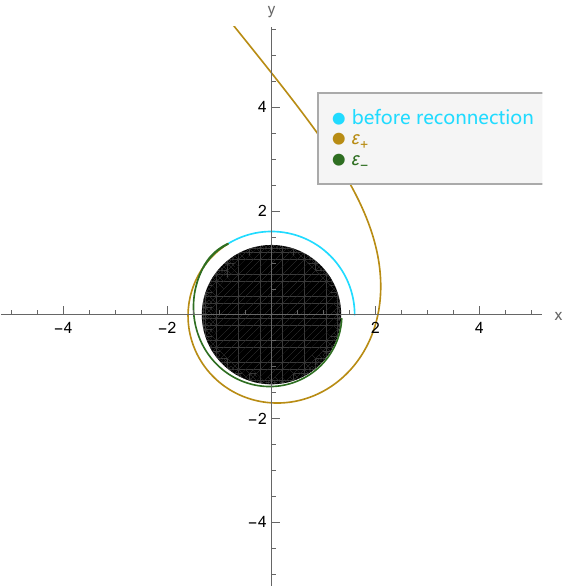}
    \caption{}
  \end{subfigure}
  \begin{subfigure}{0.24\textwidth}
    \centering
    \includegraphics[width=\linewidth]{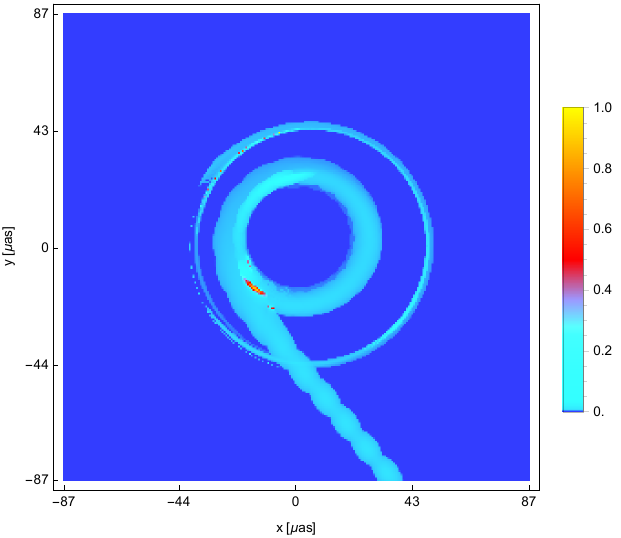}
    \caption{} 
  \end{subfigure}
  \begin{subfigure}{0.24\textwidth}
    \centering
    \includegraphics[width=\linewidth]{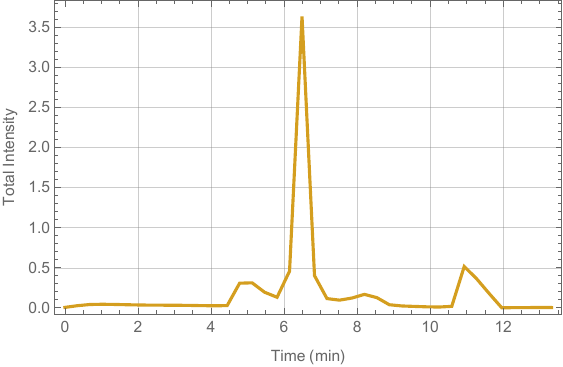}
    \caption{}
  \end{subfigure}
  \begin{subfigure}{0.24\textwidth}
    \centering
    \includegraphics[width=\linewidth]{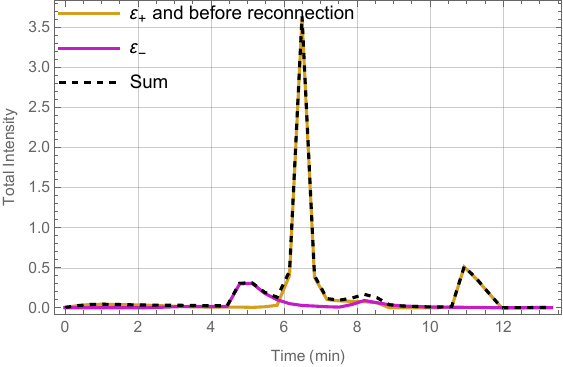} 
    \caption{}  
  \end{subfigure}
\caption{$\Lambda=-0.0002$. (a) Plasmoid trajectory in a 2D Cartesian coordinate system.    (b) Normalized intensity distribution of the observed hotspot.             (c) Emission light curve from the hotspot.              (d) Light curves under the three different scenarios.}
  \label{fig:3}
\end{figure}

\begin{figure}[!h]
  \centering
  \begin{subfigure}{0.24\textwidth}
    \centering
    \includegraphics[width=\linewidth]{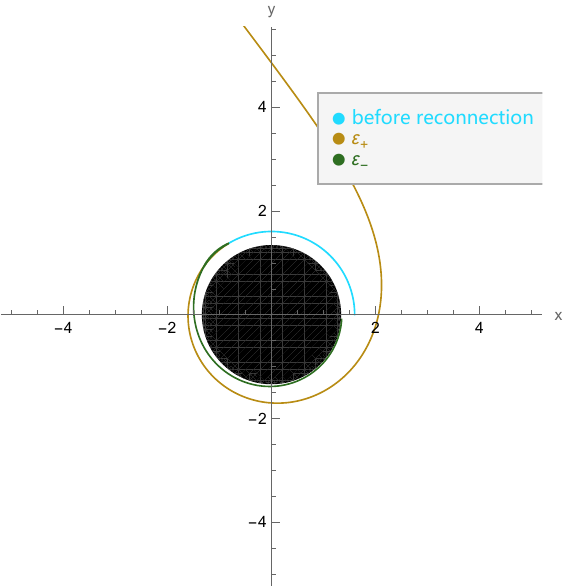}
    \caption{}
  \end{subfigure}
  \begin{subfigure}{0.24\textwidth}
    \centering
    \includegraphics[width=\linewidth]{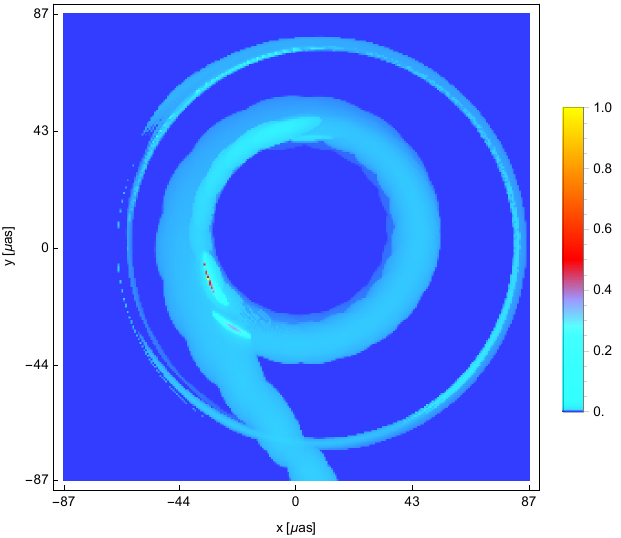}
    \caption{} 
  \end{subfigure}
  \begin{subfigure}{0.24\textwidth}
    \centering
    \includegraphics[width=\linewidth]{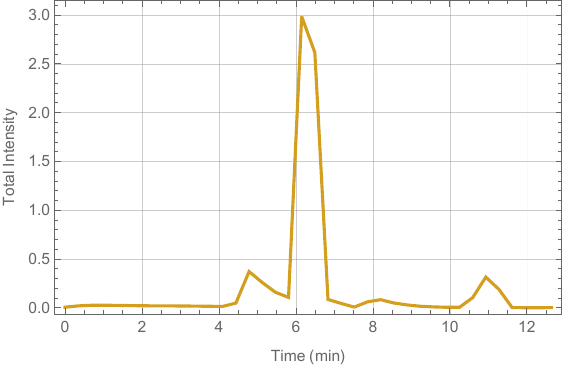}
    \caption{}
  \end{subfigure}
  \begin{subfigure}{0.24\textwidth}
    \centering
    \includegraphics[width=\linewidth]{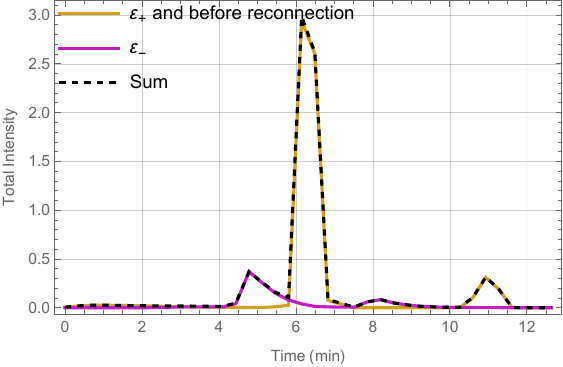} 
    \caption{}  
  \end{subfigure}
\caption{$\Lambda=-0.0007$. (a) Plasmoid trajectory in a 2D Cartesian coordinate system. (b) Normalized intensity distribution of the observed hotspot. (c) Emission light curve from the hotspot. (d) Light curves under the three different scenarios.}
  \label{fig:4}
\end{figure}

\begin{figure}[!h]
  \centering
  \begin{subfigure}{0.24\textwidth}
    \centering
    \includegraphics[width=\linewidth]{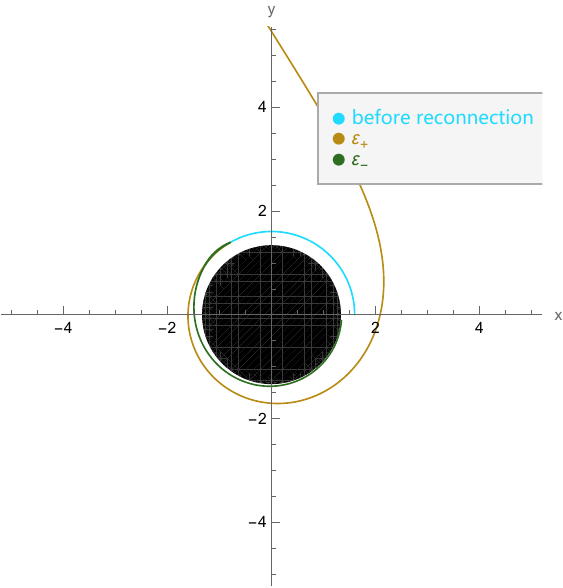}
    \caption{}
  \end{subfigure}
  \begin{subfigure}{0.24\textwidth}
    \centering
    \includegraphics[width=\linewidth]{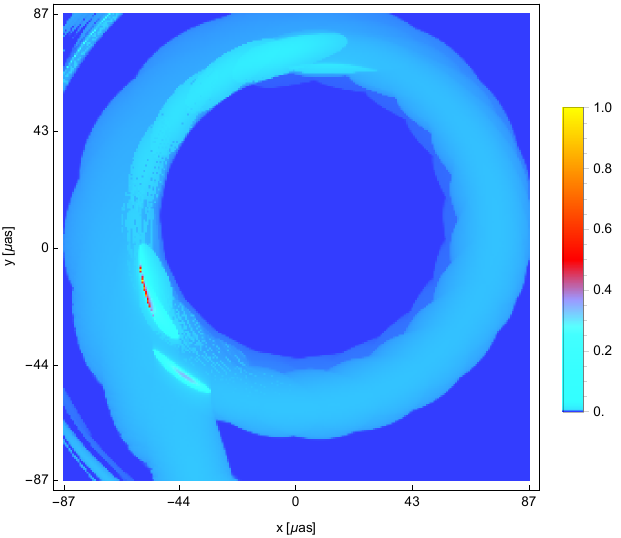}
    \caption{} 
  \end{subfigure}
  \begin{subfigure}{0.24\textwidth}
    \centering
    \includegraphics[width=\linewidth]{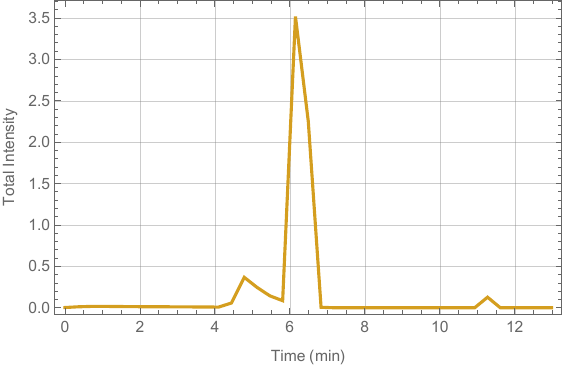}
    \caption{}
  \end{subfigure}
  \begin{subfigure}{0.24\textwidth}
    \centering
    \includegraphics[width=\linewidth]{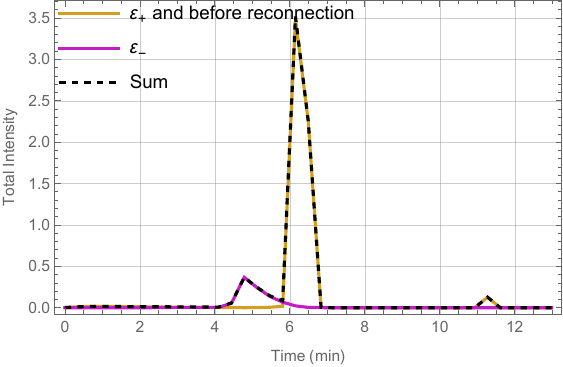} 
    \caption{}  
  \end{subfigure}
\caption{$\Lambda=-0.002$. (a) Plasmoid trajectory in a 2D Cartesian coordinate system. (b) Normalized intensity distribution of the observed hotspot. (c) Emission light curve from the hotspot. (d) Light curves under the three different scenarios.}
  \label{fig:5}
\end{figure}

Next, we vary the cosmological constant to assess its impact on hotspot imaging for different values of $\Lambda$. For $\Lambda=-0.0002$, we have $\varepsilon_{-}=-0.50334, \varepsilon_{+}=9.88827$; for $\Lambda=-0.0007$, $\varepsilon_{-}=-0.50272, \varepsilon_{+}=9.88080$; and for $\Lambda=-0.002$, $\varepsilon_{-}=-0.50100, \varepsilon_{+}=9.86241$. In all these cases, the energy extraction criteria are satisfied. The corresponding observational results are shown in Figs. \ref{fig:3}, \ref{fig:4}, and \ref{fig:5}.

From Figs. \ref{fig:2}, \ref{fig:3}, \ref{fig:4}, and \ref{fig:5}, it can be seen that after increasing the absolute value of the cosmological constant, the light curves still produce three flares. Among them, the first flare comes from $\varepsilon_{-}$, and the subsequent two flares come from $\varepsilon_{+}$, which is consistent with the previous conclusion. The intensity of the first flare is similar in all four cases, indicating that identifying the energy extraction signal is largely independent of the cosmological constant. The difference is that as the absolute value of the cosmological constant increases, the hotspot observed becomes larger, and thus the image in the normalized intensity map of the hotspot also enlarges. This differs from the previous conclusion and indicates that the presence of the cosmological constant causes the hotspot to appear larger.

\begin{figure}[!h]
  \centering
  \begin{subfigure}{0.24\textwidth}
    \centering
    \includegraphics[width=\linewidth]{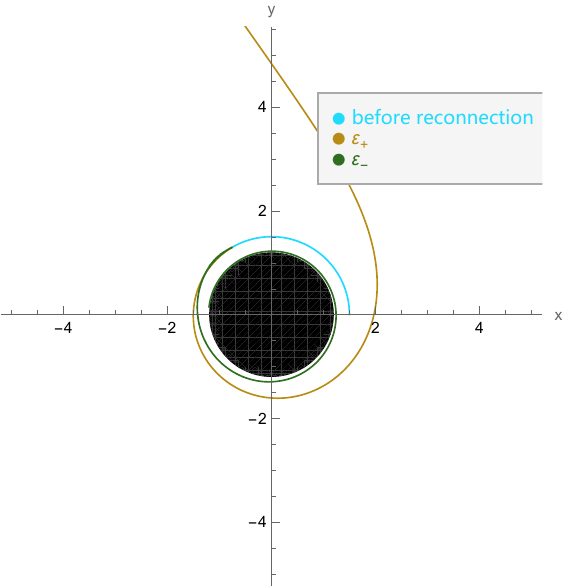}
    \caption{}
  \end{subfigure}
  \begin{subfigure}{0.24\textwidth}
    \centering
    \includegraphics[width=\linewidth]{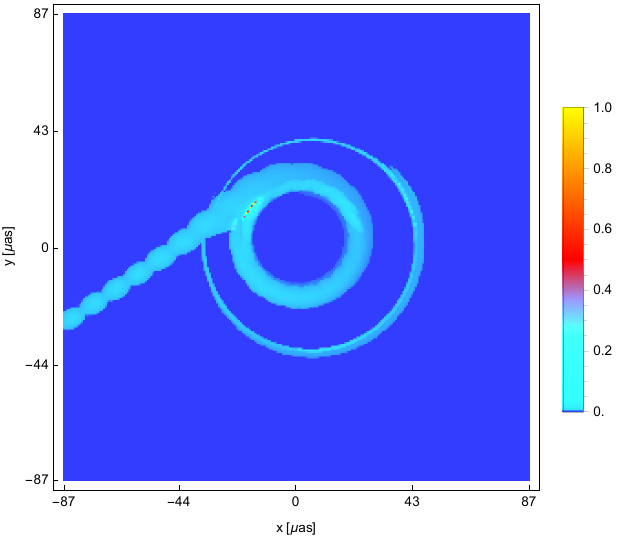}
    \caption{} 
  \end{subfigure}
  \begin{subfigure}{0.24\textwidth}
    \centering
    \includegraphics[width=\linewidth]{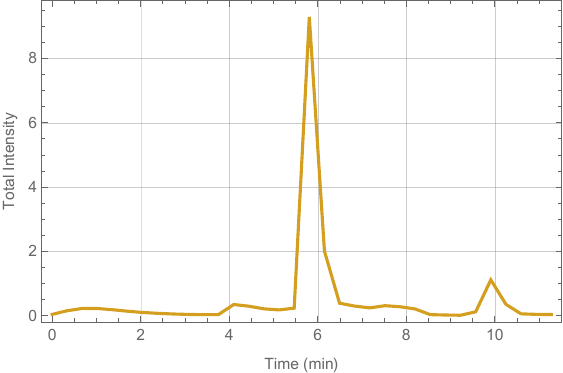}
    \caption{}
  \end{subfigure}
  \begin{subfigure}{0.24\textwidth}
    \centering
    \includegraphics[width=\linewidth]{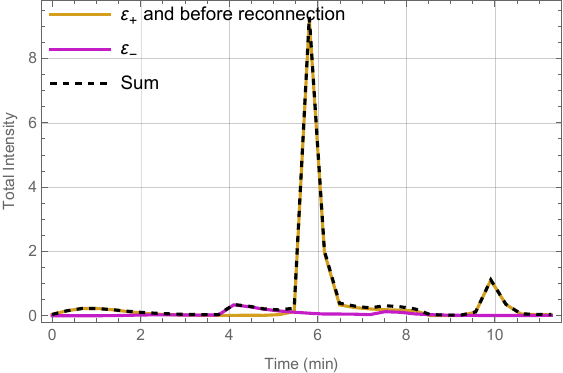} 
    \caption{}  
  \end{subfigure}
\caption{$\Lambda=-0.0001$. (a) Plasmoid trajectory in a 2D Cartesian coordinate system. (b) Normalized intensity distribution of the observed hotspot. (c) Emission light curve from the hotspot. (d) Light curves under the three different scenarios.}
  \label{fig:6}
\end{figure}

\begin{figure}[!h]
  \centering
  \begin{subfigure}{0.24\textwidth}
    \centering
    \includegraphics[width=\linewidth]{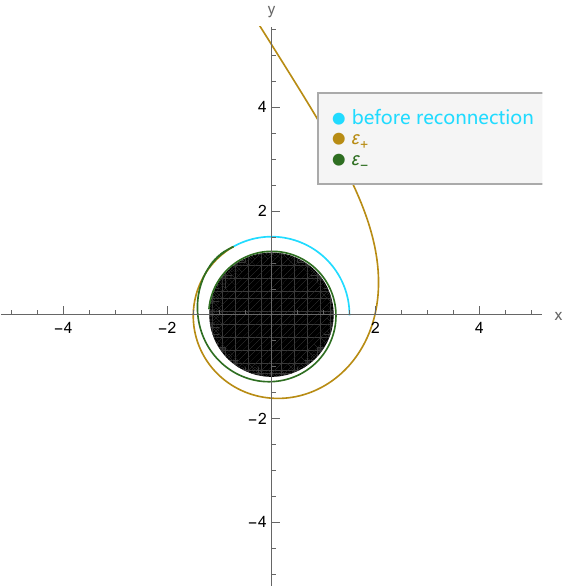}
    \caption{}
  \end{subfigure}
  \begin{subfigure}{0.24\textwidth}
    \centering
    \includegraphics[width=\linewidth]{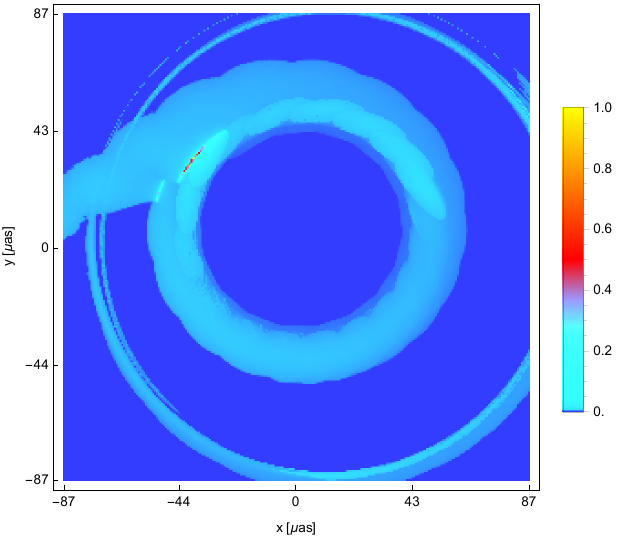}
    \caption{} 
  \end{subfigure}
  \begin{subfigure}{0.24\textwidth}
    \centering
    \includegraphics[width=\linewidth]{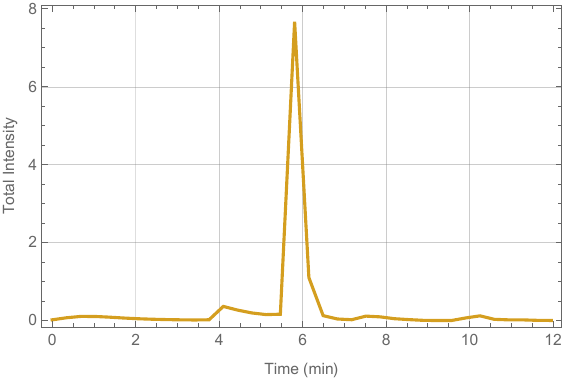}
    \caption{}
  \end{subfigure}
  \begin{subfigure}{0.24\textwidth}
    \centering
    \includegraphics[width=\linewidth]{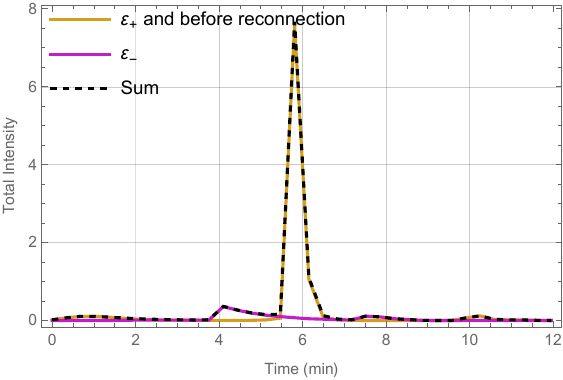} 
    \caption{}  
  \end{subfigure}
\caption{$\Lambda=-0.001$. (a) Plasmoid trajectory in a 2D Cartesian coordinate system. (b) Normalized intensity distribution of the observed hotspot. (c) Emission light curve from the hotspot. (d) Light curves under the three different scenarios.}
  \label{fig:7}
\end{figure}

Next, to verify the regularity we discovered, namely that a larger absolute value of the cosmological constant results in a larger hotspot, we change the parameters. We set $\sigma=20$, $\xi=\pi/16$, $a=0.98$, $s=0.18$, the observer's azimuthal angle $\phi_0=\pi$, the observer's inclination angle $\theta_0=\pi/10$, the observer's radial distance $r_{\text{obs}}=150$, and the radial distance of the dominant $X$-point $r_X=1.5$. For the cosmological constant, we take $\Lambda=-0.0001$ and $\Lambda=-0.001$. For these cases, $\varepsilon_{-}=-0.71153, -0.70965$ and $\varepsilon_{+}=7.77441, 7.77439$, respectively. The observational results are shown in Figs. \ref{fig:6} and \ref{fig:7}.

From Figs. \ref{fig:6} and \ref{fig:7}, it can be seen that even after changing the parameters, the trend of the hotspot image enlarging with an increase in the absolute value of $\Lambda$ still holds, which further verifies the correctness of this conclusion. These two figures also show that the intensity of the first flare is essentially the same for both values of $\Lambda$, again indicating that identifying the energy extraction signal has little dependence on the cosmological constant. Furthermore, for $\Lambda=-0.001$, the third flare is less pronounced, or in other words, the light curve effectively shows only two flares. This is somewhat similar to the case with $\Lambda=-0.002$ under different parameters, potentially suggesting that when the absolute value of $\Lambda$ is relatively large, the flare produced by the secondary image of $\varepsilon_{+}$ becomes less distinct. We attribute this to the fact that for larger absolute values of $\Lambda$, the secondary image has partially moved beyond the screen, thereby rendering the third flare less intense, as it originates from this secondary image.

Finally, we present the hotspot image for a Kerr black hole for comparison. We adopt the same parameters as in Figs. \ref{fig:6} and \ref{fig:7}, setting the cosmological constant to $\Lambda=0$. For this case, $\varepsilon_{-}=-0.71174$ and $\varepsilon_{+}=7.77442$.
\begin{figure}[!h]
  \centering
  \begin{subfigure}{0.24\textwidth}
    \centering
    \includegraphics[width=\linewidth]{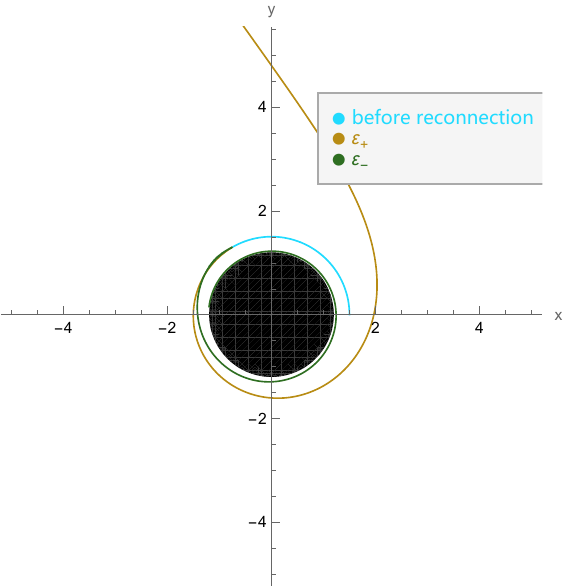}
    \caption{}
  \end{subfigure}
  \begin{subfigure}{0.24\textwidth}
    \centering
    \includegraphics[width=\linewidth]{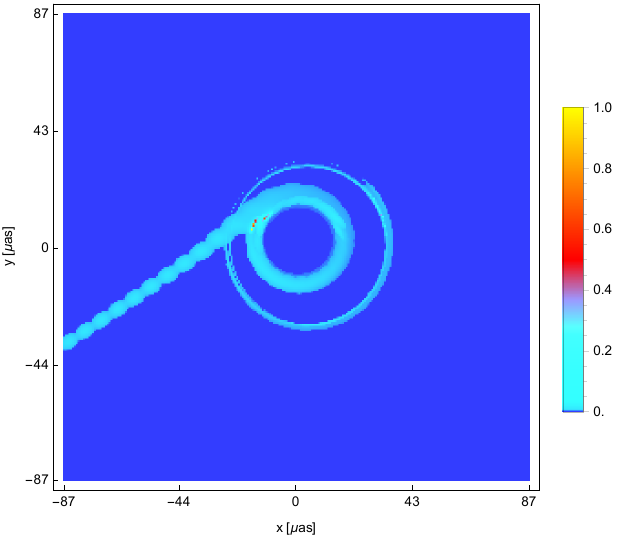}
    \caption{} 
  \end{subfigure}
  \begin{subfigure}{0.24\textwidth}
    \centering
    \includegraphics[width=\linewidth]{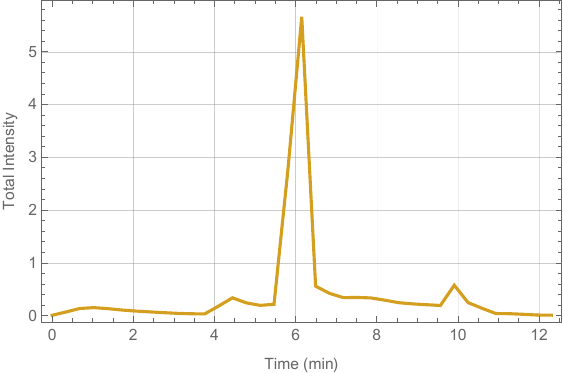}
    \caption{}
  \end{subfigure}
  \begin{subfigure}{0.24\textwidth}
    \centering
    \includegraphics[width=\linewidth]{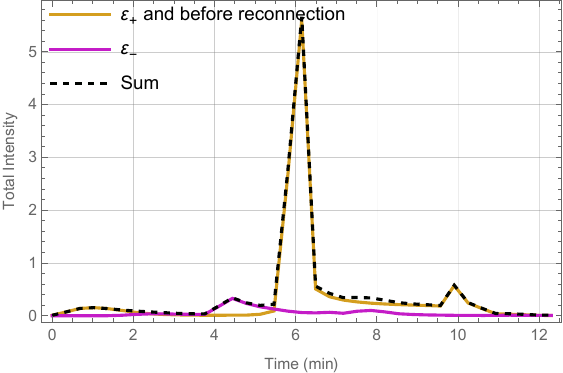} 
    \caption{}  
  \end{subfigure}
\caption{$\Lambda=0$. (a) Plasmoid trajectory in a 2D Cartesian coordinate system. (b) Normalized intensity distribution of the observed hotspot. (c) Emission light curve from the hotspot. (d) Light curves under the three different scenarios.}
  \label{fig:8}
\end{figure}
From Fig. \ref{fig:8}, it can be observed that for $\Lambda=0$, the hotspot image is smaller than that for $\Lambda=-0.0001$, which again indicates that the hotspot image enlarges as the absolute value of $\Lambda$ increases.

From Fig. \ref{fig:8}, it can be observed that for $\Lambda=0$, the hotspot image is smaller than that for $\Lambda=-0.0001$, which again indicates that the hotspot image enlarges as the absolute value of $\Lambda$ increases.

\section{Conclusion}
In this paper, we employed the hotspot imaging method within Kerr-AdS black holes to observe the motion of plasmoids following the Comisso-Asenjo process (and briefly before its onset). First, we provided a concise review of the Comisso-Asenjo process in Kerr-AdS black holes. Then, we introduced the hotspot model and the imaging methodology. Building upon these theoretical foundations, we presented the observational results obtained through numerical methods. We plotted the temporal evolution of the plasma hotspot distribution, the trajectory of the plasmoid in a two-dimensional Cartesian coordinate system, the normalized intensity distribution of the hotspot as seen by an observer, the total light curve of the hotspot emission, the light curve for $\varepsilon_{+}$ and the case without magnetic reconnection, the light curve produced solely by $\varepsilon_{-}$, and the normalized hotspot intensity distribution produced solely by $\varepsilon_{-}$. We categorized our observational results based on two parameter sets, where the first set used $\sigma=25$, $\xi=\pi/20$, $a=0.94$, $s=0.2$, observer azimuth $\phi_0=\pi/2$, observer inclination $\theta_0=\pi/10$, observer radial distance $r_{\text{obs}}=200$, and dominant $X$-point radial distance $r_X=1.6$, while the second set used $\sigma=20$, $\xi=\pi/16$, $a=0.98$, $s=0.18$, $\phi_0=\pi$, $\theta_0=\pi/10$, $r_{\text{obs}}=150$, and $r_X=1.5$. Through analysis of different cosmological constants within these parameter configurations, we reached the significant conclusion that the hotspot image enlarges with increasing absolute value of $\Lambda$, demonstrating the notable influence of the cosmological constant on hotspot characteristics.

This study focused exclusively on a negative cosmological constant (AdS spacetime). Investigating the effects of a positive cosmological constant (dS spacetime) is a promising avenue for future work. Since black hole event horizons, ergospheres, and magnetic reconnection can also exist in such spacetimes, exploring whether similar trends hold there is a compelling research question.

\noindent {\bf Acknowledgments}

\noindent
This work is supported by the National Natural Science Foundation of China (Grants Nos. 12375043,
12575069 ).


\begin{thebibliography}{99}

\bibitem{5}
K.~Akiyama \textit{et al.} [Event Horizon Telescope],
``First M87 Event Horizon Telescope Results. I. The Shadow of the Supermassive Black Hole,''
Astrophys. J. Lett. \textbf{875} (2019), L1
\bibitem{6}
K.~Akiyama \textit{et al.} [Event Horizon Telescope],
``First Sagittarius A* Event Horizon Telescope Results. I. The Shadow of the Supermassive Black Hole in the Center of the Milky Way,''
Astrophys. J. Lett. \textbf{930} (2022) no.2, L12
\bibitem{7}
R.~Abuter \textit{et al.} [GRAVITY],
``Polarimetry and astrometry of NIR flares as event horizon scale, dynamical probes for the mass of Sgr A*,''
Astron. Astrophys. \textbf{677} (2023), L10
\bibitem{8}
R.~Abuter, A.~Amorim, M.~Baub{\"o}ck, J.~P.~Berger, H.~Bonnet, W.~Brandner, Y.~Cl{\'e}net, V.~Coud{\'e} du Foresto, P.~T.~de Zeeuw and C.~Deen, \textit{et al.}
``Detection of orbital motions near the last stable circular orbit of the massive black hole SgrA*,''
Astron. Astrophys. \textbf{618} (2018), L10
\bibitem{12}
L.~Meyer, A.~Eckart, R.~Schoedel, W.~J.~Duschl, K.~Muzic, M.~Dovciak and V.~Karas,
``Near-infrared polarimetry setting constraints on the orbiting spot model for Sgr A* flares,''
Astron. Astrophys. \textbf{460} (2006), 15
\bibitem{13}
S.~Trippe, T.~Paumard, T.~Ott, S.~Gillessen, F.~Eisenhauer, F.~Martins and R.~Genzel,
``A polarised infrared flare from Sagittarius A* and the signatures of orbiting plasma hotspots,''
Mon. Not. Roy. Astron. Soc. \textbf{375} (2007), 764-772
\bibitem{14}
N.~Hamaus, T.~Paumard, T.~Muller, S.~Gillessen, F.~Eisenhauer, S.~Trippe and R.~Genzel,
``Prospects for testing the nature of Sgr A*'s NIR flares on the basis of current VLT- and future VLTI-observations,''
Astrophys. J. \textbf{692} (2009), 902-916
\bibitem{15}
V.~I.~Dokuchaev and N.~O.~Nazarova,
``Modeling the motion of a bright spot in jets from black holes M87* and SgrA*,''
Gen. Rel. Grav. \textbf{53} (2021) no.8, 83
\bibitem{16}
M.~Baub{\"o}ck \textit{et al.} [GRAVITY],
``Modeling the orbital motion of Sgr A*{\textquoteright}s near-infrared flares,''
Astron. Astrophys. \textbf{635} (2020), A143
\bibitem{17}
Y.~Chen, P.~Wang and H.~Yang,
``Interferometric signatures of black holes with multiple photon spheres,''
Phys. Rev. D \textbf{110} (2024) no.4, 044020
\bibitem{18}
J.~L.~Rosa, D.~S.~J.~Cordeiro, C.~F.~B.~Macedo and F.~S.~N.~Lobo,
``Observational imprints of gravastars from accretion disks and hot spots,''
Phys. Rev. D \textbf{109} (2024) no.8, 084002
\bibitem{19}
F.~A.~Asenjo and L.~Comisso,
``Relativistic Magnetic Reconnection in Kerr Spacetime,''
Phys. Rev. Lett. \textbf{118} (2017) no.5, 055101
\bibitem{20}
L.~Comisso and F.~A.~Asenjo,
``Collisionless Magnetic Reconnection in Curved Spacetime and the Effect of Black Hole Rotation,''
Phys. Rev. D \textbf{97} (2018) no.4, 043007
\bibitem{21}
Z.~Y.~Fan, Y.~Li, F.~Zhou and M.~Guo,
``Fast magnetic reconnection in Kerr spacetime,''
Phys. Rev. D \textbf{110} (2024) no.10, 104044
\bibitem{22}
Z.~Y.~Fan, F.~Zhou, Y.~Li, M.~Guo and B.~Chen,
``Magnetic reconnection under centrifugal and gravitational electromotive forces,''
Phys. Rev. D \textbf{111} (2025) no.6, 064067 
\bibitem{23}
R.~Penrose,
``Gravitational collapse: The role of general relativity,''
Riv. Nuovo Cim. \textbf{1} (1969), 252-276
\bibitem{1}
L.~Comisso and F.~A.~Asenjo,
``Magnetic Reconnection as a Mechanism for Energy Extraction from Rotating Black Holes,''
Phys. Rev. D \textbf{103} (2021) no.2, 023014
\bibitem{24}
X.~X.~Zeng and K.~Wang,
``Energy extraction from the Kerr-Bertotti-Robinson black hole via magnetic reconnection in a circular and a plunging plasma,''
Phys. Rev. D \textbf{112} (2025) no.6, 064032
\bibitem{25}
K.~Wang and X.~X.~Zeng,
``Energy extraction from the accelerating Kerr black hole via magnetic reconnection in the plunging region and circular orbit region,''
[arXiv:2508.11934 [gr-qc]].
\bibitem{26}
X.~X.~Zeng and K.~Wang,
``Energy extraction via magnetic reconnection in Kerr-Sen-AdS4 black hole: Circular plasma and plunging plasma,''
Phys. Rev. D \textbf{112} (2025) no.6, 064080
\bibitem{27}
S.~J.~Zhang,
``Energy extraction via magnetic reconnection in Konoplya-Rezzolla-Zhidenko parametrized black holes,''
Phys. Rev. D \textbf{109} (2024) no.8, 084066
\bibitem{28}
S.~Shaymatov, M.~Alloqulov, B.~Ahmedov and A.~Wang,
``Kerr-Newman-modified-gravity black hole{\textquoteright}s impact on the magnetic reconnection,''
Phys. Rev. D \textbf{110} (2024) no.4, 044005
\bibitem{29}
F.~Long, S.~Wang, S.~Chen and J.~Jing,
``Magnetic reconnection and energy extraction from a Konoplya{\textendash}Zhidenko rotating non-Kerr black hole,''
Eur. Phys. J. C \textbf{85} (2025) no.1, 26
\bibitem{38}
S.~Shaymatov, M.~Alloqulov, B.~Ahmedov and A.~Wang,
``Kerr-Newman-modified-gravity black hole{\textquoteright}s impact on the magnetic reconnection,''
Phys. Rev. D \textbf{110} (2024) no.4, 044005
\bibitem{33}
Z.~Zhao, Z.~Y.~Fan, X.~Wang, M.~Guo and B.~Chen,
``Probing the Penrose Process: Images of Split Hotspots and Their Observational Signatures,''
[arXiv:2510.27409 [astro-ph.HE]].
\bibitem{34}
K.~Wang and X.~X.~Zeng,
``Hotspot Images Driven by Magnetic Reconnection in Kerr-Sen black hole,''
[arXiv:2511.01342 [gr-qc]].
\bibitem{30}
S.~W.~Hawking and G.~F.~R.~Ellis,
``The Large Scale Structure of Space-Time,''
(Cambridge University Press, Cambridge, England, 1973)
\bibitem{31}
J.~M.~Maldacena,
``The Large $N$ limit of superconformal field theories and supergravity,''
Adv. Theor. Math. Phys. \textbf{2} (1998), 231-252
\bibitem{32}
E.~Witten,
``Anti de Sitter space and holography,''
Adv. Theor. Math. Phys. \textbf{2} (1998), 253-291


\bibitem{2}
B.~Carter,
``The commutation property of a stationary, axisymmetric system,''
Commun. Math. Phys. \textbf{17} (1970), 233-238
\bibitem{3}
Z.~Stuchlik and P.~Slany,
``Equatorial circular orbits in the Kerr-de Sitter space-times,''
Phys. Rev. D \textbf{69} (2004), 064001
\bibitem{4}
P.~Slany and Z.~Stuchlik,
``Relativistic thick discs in the Kerr-de Sitter backgrounds,''
Class. Quant. Grav. \textbf{22} (2005), 3623-3652
\bibitem{9}
Z.~Hu, Z.~Zhong, P.~C.~Li, M.~Guo and B.~Chen,
``QED effect on a black hole shadow,''
Phys. Rev. D \textbf{103} (2021) no.4, 044057
\bibitem{10}
Y.~Hou, Z.~Zhang, H.~Yan, M.~Guo and B.~Chen,
``Image of a Kerr-Melvin black hole with a thin accretion disk,''
Phys. Rev. D \textbf{106} (2022) no.6, 064058





\bibitem{Zeng:2025kqw}
X.~X.~Zeng, C.~Y.~Yang, M.~I.~Aslam, R.~Saleem and S.~Aslam,
``Kerr-like black hole surrounded by cold dark matter halo: the shadow images and EHT constraints,''
JCAP \textbf{08} (2025), 066


\bibitem{Zeng:2021dlj}
X.~X.~Zeng, G.~P.~Li and K.~J.~He,
``The shadows and observational appearance of a noncommutative black hole surrounded by various profiles of accretions,''
Nucl. Phys. B \textbf{974} (2022), 115639


\bibitem{Zeng:2021mok}
X.~X.~Zeng, K.~J.~He and G.~P.~Li,
``Effects of dark matter on shadows and rings of Brane-World black holes illuminated by various accretions,''
Sci. China Phys. Mech. Astron. \textbf{65} (2022) no.9, 290411










\bibitem{11}
J.~Huang, Z.~Zhang, M.~Guo and B.~Chen,
``Images and flares of geodesic hot spots around a Kerr black hole,''
Phys. Rev. D \textbf{109} (2024) no.12, 124062

\end{thebibliography}
\end{document}